\def\nref  {\noindent\parshape 2 0.0 truein 06.5 truein 0.4 truein 06.1 truein}

\def\\{$\backslash$}
\def\deg      {{\ifmmode^\circ\else$^\circ$\fi} } 
\def\arcm     {{\ifmmode {'\ }\else$'     $\fi} } 
\def\arcs     {{\ifmmode{''\ }\else$''    $\fi} } 

\def\cge      {{$_ >\atop{^\sim}$}}

\def\Lsun     {{$L_{\odot}$} }

\def\Msun     {{\ $M_{\odot}$} }

\def\nref  {\noindent\parshape 2 0.0 truein 06.5 truein 0.4 truein 06.1 truein}

\def\nref  {\noindent\parshape 2 0.0 truein 06.5 truein 0.4 truein 06.1 truein}
\def\Msun     {{\ $M_{\odot}$} }

\documentstyle[12pt,aasms4,flushrt]{article}

\begin{document}

\title{\bf Radio Identification of Sub-mm Sources in the Hubble Deep Field}

\author{E. A. Richards}

\affil{National Radio Astronomy Observatory\footnotemark[1] and University of Virginia}
\footnotetext[1]{The National Radio Astronomy Observatory is a facility
of the National
Science Foundation operated under cooperative agreement by Associated
Universities, Inc.}
\authoraddr{
 520 Edgemont Road, Charlottesville, Virginia 22903 \newline
Electronic mail: er4n@virginia.edu}

\begin{abstract}

        Determination of the epoch dependent star-formation
rate of field galaxies is one of the principal goals
of modern observational cosmology.
Recently, Hughes et al. (1998) using the SCUBA
instrument on the James Clerk Maxwell Telescope,
report the detection of a new population of heavily
dust enshrouded, star-forming galaxies at high
redshifts ($z >$2), dramatically altering the picture
of galaxy evolution. However, we show that this
interpretation
must be treated with caution because of
ambiguities in the identification of the host
galaxies. Based on our deep, high resolution 1.4 GHz
obervations of the Hubble Deep Field, we suggest
alternate identifications to the sub-mm detections.
These identifications 
argue for a substantially lower redshift to the
sub-mm population with a consequential lowering
of the $z>$2 sub-mm/far infrared luminosity density
and global star-formation rate.

\end{abstract}

\keywords{cosmology: observations --- galaxies: evolution --- 
galaxies: starburst, radio continuum}

\slugcomment{Submitted to {\it The Astrophysical Journal Letters}, November
1998}

\section{Introduction}

        The data used to make a quantitative
estimate of
the global star-formation history typically
 consist of optical multi-color imaging combined with
comprehensive spectroscopy (Lilly et al. 1996).
These data are then used to infer the distances
of individual galaxies and (with the aid
of stellar synthesis models) determine
their intrinsic spectral energy distributions.
In this manner active star-formation rates
can be calculated (Madau et al. 1996, Connelly et al. 1997).
The primary limitation of this technique
is its sensitivity to dust obscuration, the effects of
which in evolving galaxy populations
are both uncertain and controversial (Madau et al. 1996, Calzetti
1997, Meuer et al. 1997).

        An alternative method
for the study of star-forming galaxies
is to observe directly the reprocessed UV light
emitted by the dust. Depending on the
dust temperature (30 K - 60 K), this emission peaks
between 50-100 $\mu$m (restframe) and in many galaxies
comprises the bulk of the bolometric luminosity. Additionally,
the thermal far infrared radiation (FIR)
is not subject to further obscuration, so uncertain
extinction corrections are avoided. At high
redshifts, the FIR emission will be
shifted into the sub-mm band.
A consequence of the steep Rayleigh-Jeans
tail of the thermal dust emission coupled with a
positive dust emissivity index, $\alpha$ (where the emissivity,
$\epsilon \propto \nu ^{\alpha}$)
 is to bring higher
flux into the sub-mm band so the cosmological
dimming effect is offset. Thus a starburst galaxy of a given
luminosity should have essentially the same observed flux
density between  $z$ = 1 and 10 (Blain et al. 1993).

        Closely related to the FIR emission
in starburst galaxies is the radio continuum.
In normal galaxies (i.e., without a powerful AGN),
the centimeter radio luminosity is dominated
by diffuse synchrotron emission believed to be
produced by relativistic electrons accelerated
in supernovae remnants.
At shorter wavelengths free-free
emission from HII regions may contribute
substantially. Although the radio emission
is linked to active star-formation by
different physical mechanisms than that
of the FIR, there is a tight correlation
between the FIR and radio luminosity of
a starburst (Helou et al. 1985, Condon et al. 1991).
Radio observations are only sensitive to
recent starburst activity in a galaxy
(and the formation of its O and B
stellar populations) since the thermal and
synchrotron radiation dissipate on physical
time scales of $10^7-10^8$ yr. In
this sense the radio luminosity of a
starburst is a true measure of the
instantaneous rate of star-formation
in a galaxy, uncontaminated by older stellar
populations. Because galaxies and the inter-galactic
medium are transparant at centimeter wavelengths, 
radio emission is a sensitive measure of
star-formation in distant galaxies.
The current deep Very
Large Array (VLA) radio surveys
with sensitivites of a few microjansky are
capable of detecting star-forming galaxies
to $z\sim$1.5 (Richards et al. 1998).

        An 850 $\mu$m survey of the HDF with the SCUBA detector
on the James Clerk
Maxwell Telescope (JCMT) detected five sources
in a confusion limited image (Hughes et al. 1998)
with 15\arcsec ~resolution.
Tentative optical identifications
are all with putative starbursts at $z$ \cge 1
and with star-formation rates (SFR)
of 400-1200\Msun yr$^{-1}$ (we assume $h$ = 0.5, $q_o$ = 0.5)
In this letter we 
compare our deep radio images of the Hubble
Deep Field with the SCUBA images and suggest alternate
optical counterparts to the sub-mm sources.

\section{Radio Observations}

       A 5.4\arcmin ~(FWHM) field containing the HDF
has been imaged at 8.5 GHz with the VLA
with an rms sensitivity of 1.8 $\mu$Jy. The observing
technique and data reduction are discussed in Richards
et al. (1998). We collected 40 additional hours
of data at 8.5 GHz in June 1997. The new combined image
has a rms sensitivity near the field center of about 
1.6 $\mu$Jy with a resolution of 3.5\arcsec .

	During November 1996, we obtained
42 hours of VLA data in its A-array on the HDF 
at 1.4 GHz. The subsequent 1.4
GHz image of the HDF has an effective resolution
of 1.9\arcsec ~ and an rms sensitivity
of 7.5 $\mu$Jy with a firm detection limit of 
40 $\mu$Jy. A total of 381 radio sources at 1.4 GHz
have been catalogued within 20\arcmin ~of the HDF and 
are reported on elsewhere (Richards 1998).
Within the Hubble Deep Field, there are
nine radio sources detected in complete samples
at either 1.4 GHz and/or 8.5 GHz, while an additional
seven 8.5 GHz sources comprise a supplementary sample
as described by Richards et al. (1998). The 8.5 and
1.4 GHz images of the HDF are available at the
world-wide web site: www.cv.nrao.edu/~jkempner/vla-hdf/.

\section{Association of Sub-mm and Radio Sources}

We inspected our radio images to determine
if any of the SCUBA sources have possible radio
counterparts. Our first step was to align the
radio and SCUBA position frames.
The VLA coordinate grid
is within 0.1\arcsec ~of the J2000/FK5 reference frame
at both 8.5 and 1.4 GHz (Richards et al. 1998).
In order to tie the JCMT coordinate grid to
this system, we have assumed
that the radio and sub-mm sources are
associated with the same galaxy, as discussed
in the introduction. 

	The relative rms positional errors
for the sub-mm sources should be of order 1-2\arcsec ~
(based on the 15\arcsec ~SCUBA beam size
and the signal to noise ratio of individual
detections), howver the uncertain effects of 
source confusion in the sub-mm images likely makes 
this an underestimate. In addition, the 
absolute registration of the SCUBA image
is unknown a priori, although Hughes et al.
(1998) quote a value of 0.5\arcsec ,
while Smail et al. (1998) report typical values
of 3\arcsec ~for their SCUBA images.
Thus we chose to search for any radio object either
in the 1.4 GHz complete sample or in the 8.5 GHz
 catalog of Richards et al. (1998) within a
10\arcsec ~error circle around each of the
sub-mm source positions.
Possible radio associations
were apparent for HDF850.1 (3651+1226 and 3651+1221), HDF850.2 
(3656+1207), and HDF850.4 (3649+1313). Two of these
radio sources were detected at both 1.4 and 8.5 GHz as
part of independent and complete samples (3651+1221 and
3649+1313). These two radio sources in particular 
also have high signifigance ISO 15 $\mu$m counterparts 
(Aussel et al. 1998) indicating these systems 
may contain substantial amounts of dust and hence
be luminous FIR galaxies. Based on the
association of 3651+1221 and 3649+1313
with HDF850.1 and HDF850.4, respectively, we suggest
a SCUBA coordinate frame shift of 4.8\arcsec 
~west and 3.8\arcsec ~south.

	The position shifts, within
1.2\arcsec ~in the (VLA+ISO) vs. SCUBA positions
for both HDF850.1 and HDF850.4, suggests that the
VLA/ISO/SCUBA source alignment is not accidental.
However this large registration offset of $\sim$6\arcsec~
is much greater than the 0.5\arcsec ~registraion accuracy
quoted in Hughes et al. (1998). Either the radio/ISO emission
is not associated withe the same galaxies as the SCUBA sources,
or the SCUBA observations have a large registration error.

\section{Optical Identification of Radio/Sub-mm Sources}

Since the radio source positions are much
more accurate (0.1-0.2\arcsec ) than the sub-mm positions
and as the HDF contains a high surface density of optical
objects (typically 20 per SCUBA beam), we now use the radio
data to make the secure identifications with optical
counterparts. 
Table 1 presents plausible radio counterparts
to the sub-mm sources of Hughes et al. (1998).
The first line gives the position of the SCUBA source
after translation to the radio coordinate frame
with plausible radio counterparts and their
suggested optical identification given in following
lines.

{\bf HDF850.1 :} We present in Figure 1, the 1.4 GHz
overlay of the Hubble Deep Field  centered on the 
shifted SCUBA position.
The precise 1.4 GHz radio position 
suggests the optical identification is with the 
faint low-surface brightness
object to the immediate north of
the brighter foreground disk system. Richards et al.
(1998) suggested that this optical feature might be
associated with the $z = 0.299$ galaxy 3-659.1.
However, we note the presence of a separate low 
surface brightness galaxy at $z$ = 1.72 
(3-633.1; Fernandez-Soto et al. 1998) located approximately
2\arcsec ~to the northwest. 

If radio source 3651+1221 is the  most obscured part of
a larger galaxy 3-633.1 at $z$ = 1.72 (Fernandez-Soto et al.
1998), 
the implied radio luminosity (L$_{1.4}$ = 10$^{25}$ W/Hz)
is substantially higher than that of the most extreme
local starbursts (e.g., ARP 220) and suggests that 
this object may contain an AGN. 

{\bf HDF850.2 :} Based on the optical/radio positional
coincidence, this 3.5 $\sigma$ radio source
has a 90\% reliability of being associated
with a I=23.7 distorted galaxy (Barger et al. 1998),
according to
the analysis of Richards et al. (1998).
We identify the 850 $\mu$m detection with this
galaxy of unknown redshift. The $UGR$ band
photometry (Hogg 1997) on this galaxy suggests that it is likely
at $z < 3$. Figure 2 shows the 1.4 GHz overlay of the optical
field.

{\bf HDF850.3 :} The radio data does little to 
clarify the identification of the sub-mm source
in this field. Figure 3 shows a 4 $\sigma$ radio
 source located
4\arcsec ~ from the position of the 850 $\mu$m
detection. However, before the shift the 
SCUBA source position is in good agreement with
the position of the bright disk system 1-34.0,
which is also included in the supplemental ISO catalog of
Aussel et al. (1998). This galaxy also has a weak
3$\sigma$ radio detection. The 0.485 redshift of
this object implies a star-formation rate of  80 \Msun yr$^{-1}$
from the radio luminosity (Salpeter initial mass function
integrated over 0.1-100\Msun ), although the presence of an
AGN cannot be ruled out.
At present the data cannot discriminate between these 
two possible radio/sub-mm associations.

{\bf HDF850.4 :}  The  radio source 3649+1313 is
associated with the spiral galaxy 2-264.1
at a redshift of 0.475. ISO sources from the
complete catalogs of Aussel et al. (1998)
and Rowan-Robinson et al. (1997) have been
assoicated with this radio source.
This galaxy is
likely part of the larger structure at 0.475
which contains 16 galaxies with spectroscopic
redshifts (Cohen et al. 1996). At least one of
these galaxies (2-264.2) lies at a small
projected distance (less than 30 kpc) and
suggests  dynamic interactions may be triggering
the radio/sub-mm activity. Although the SCUBA
detection may be a blend of emission from
several galaxies in this crowded field (see Figure 4), the radio
emission is clearly confined to the central galaxy
2-264.1. Richards et al. (1998) estimate a 
SFR = 150\Msun yr$^{-1}$.

	Can the SCUBA source HDF850.4
plausibly be associated instead
with the HDF optical galaxy 2-399.0 as claimed by Hughes
{\em et al.}? If we take those authors' estimate of the FIR luminosity
log$_{10}$L$_{60 \mu m}$ = 12.47 \Lsun for this galaxy, the
FIR-radio relation (Condon et al. 1991) predicts an observed 1.4 GHz
flux density of about 300 $\mu$Jy, clearly in excess
of our upper limit of 23 $\mu$Jy (3 $\sigma$).
We find the identification of SCUBA source HDF850.4
with HDF galaxy 2-399 to be dubious and instead
identify HDF850.4 with 2-264.1.

{\bf HDF850.5 :} There is no 1.4 GHz radio emission apparent to 
the 2 $\sigma$ limit of 15 $\mu$Jy in this 
field. Optically there are only two plausible 
identifications for the 850 $\mu$m source, HDF galaxies 2-395.0
and 2-349.0 (see Figure 5). The redshift of 2-395.0 is 0.72 (Fernandez-Soto
et al. 1998).
The radio flux
limit on this galaxy can exclude a SFR $\geq$ 60 \Msun /yr.
The other possible identification (2-349.0) is at an 
unknown redshift. 
The non-detection of this sub-mm
source at radio or optical wavelengths coupled
with the fact that this is the weakest source in the
Hughes et al. (1998) catalog suggests 
this source may be spurious. We also note that 
this sub-mm source is located only 12$\arcsec$ ~from 
sub-mm detection HDF850.4
and hence may suffer from confusion.

\section{Conclusions}

        Of the five 850 $\mu$m sources in the HDF,
two are solidly detected at radio wavelengths, 
while two are probable detections. Two of these
identifications are possibly with $z \sim$ 0.5
starbursts (HDF850.3 and HDF850.4). The other
two detections (HDF850.1 and HDF850.2) must have
redshifts less than 1.5 or be contaminated by
AGN based on radio luminosity arguments.
This radio analysis suggests  the
claim, based on sub-mm observations alone,
that the optical surveys underestimate the
$z>2$ global star-formation rate are premature.
On the other hand, the $z < 1$ star-formation
history may have been underestimated if a significant
fraction of the sub-mm population lies at relatively
low redshift. 

	In the absence of high resolution 
sub-mm imaging capability, it is necessary
to rely on plausible radio counterparts
of sub-mm sources in order to provide the
astrometric accuracy needed to make the
proper optical identifications.
Only complete redshift samples of the sub-mm
population coupled with diagnostic spectroscopy
and high resolution radio data will allow 
for calculation of the epoch dependent
sub-mm luminosity function and its implication
for the star-formation history of the universe.

\acknowledgements

        We thank our collaborators K. Kellermann, E. Fomalont,
B. Partridge, R. Windhorst, and D. Haarsma
for a critical reading of an earlier version of this work.
We appreciate useful conversations with J. Condon 
and A. Barger. 
Support for part of
this work was provided by NASA through grant AR-6337.*-96A from the
Space Telescope Science Institute, which is operated by the Association
of Universities for Research in Astronomy, Inc., under NASA contract
NAS5-2655, and by NSF grant AST 93-20049.

\newpage

\nref Aussel, H. et al. A\& A, in press, 1998

\nref  Barger em et al. 1998, AJ, submitted

\nref  Blain, A. W. \& Longair, M. S. 1993, MNRAS, 264, 509

\nref  Calzetti, D. 1997, AJ, 113

\nref  Cohen, J. G. et al. 1996, ApJL, 471, 5

\nref  Condon, J. J., Anderson, M. L. \& Helou, G. 1991,
	ApJ, 376, 95

\nref  Connelly, A. J., Szalay, A. S., Dickinson, M., SubbaRao, M. V.,
  \& Brunner, R. J. 1997, ApJL, 486, 11

\nref  Fernandez-Soto, A., Lanzetta, K. M., \& Yahill, A. 1998,
	AJ, submitted

\nref  Helou, G., Soifer, B. T.,\& Rowan-Robinson, M. 1985,
	ApJL, 298, 11

\nref Hogg, D. 1998, Ph.D. thesis (Caltech)

\nref  Hughes et al. 1998, Nature, 393, 241

\nref  Lilly, S. J., Le Fevre,O., Hammer, F., Crampton, D. 
	1996, ApJL, 460, 1

\nref Lowenthal, J., et al. 1997, ApJ, 481, 673.

\nref Madau, P. et al. 1996, MNRAS, 283, 1388

\nref  Meurer, G. R., Heckman, T., Lehnert, M. D., Leitherer, C.
  \& Lowenthal, J.  1997, AJ, 114, 54

\nref Richards, E. A. 1998, in preparation

\nref  Richards, E. A., Kellermann, K. I., Fomalont, E. B., Windhorst, R. A.
        \& Partridge, R. B. 1998, AJ, 116, 1039

\nref Rowan-Robinson, M. et al. 1997, MNRAS, 289, 490

\nref  Williams et al. 1996, AJ, 112, 1335

\newpage

\section*{Figure Captions}

\figcaption{The greyscale shows a 20\arcsec $\times$ 20\arcsec ~HDF
I-band image (Williams et al. 1996) containing the
SCUBA detection HDF850.1. The contours correspond to
1.4 GHz emission at the -2, 2, 4 and 6 $\sigma$ level ($\sigma$ = 7.5 $\mu$Jy).
The three sigma position error circle for HDF850.1 is shown
after shifting to the VLA coordinate frame. The original
position of HDF850.1 taken from Hughes et al. (1998) is
denoted by the diamond. The ISO detection is marked with
a cross with three sigma position errors (Aussel et al. 1998).
The radio emission is clearly confined to the
optical feature north of the bright spiral. 
Radio source 3651+1221 may be the most obscured part of
a larger galaxy 3-633.1 at $z$ = 1.72 (Fernandez-Soto et al.
1998).}

\figcaption{The greyscale shows a 20\arcsec $\times$ 20\arcsec ~
ground-based I-band image taken from Barger et al. (1998)
centered on the position of SCUBA source HDF850.2. The 1.4 GHz radio
contours are drawn at -2, 2, 3 and 5 $\sigma$.
The symbols are the 
same as for Figure 1. We identify HDF850.2 with the 3.5 $\sigma$
radio source 3657+1159.}

\figcaption{The greyscale corresponds to optical I-band 
emission in the field of SCUBA source HDF850.3 (20\arcsec ~on a side)
 as taken from the HDF. Radio contours at 1.4 GHz  are drawn at the
-2, 2 and 4 $\sigma$ level. Intriguingly, there is a 4.2 $\sigma$
radio 'source' located 4\arcsec ~from HDF850.3. The probability
of this being a chance coincidence is 20\% based on the surface
density of 4$\sigma$ radio sources in the field. If 
HDF850.3 is associated with the radio source then this is
a blank field object to I$_{AB}$ = 27 (object lies in the less
sensitive PC). However, an ISO source from the supplemental
catalog of Aussel et al. (1998) is also in the field
and associated with the bright disk galaxy 1-34.0 (Williams
et al. 1996). It is difficult to discrimante between these
two possible sub-mm identifications with the present data.
The symbols are the
same as for Figure 1.
}

\figcaption{Radio 1.4 GHz contours drawn at the 
-2, 2, 4 and 6 $\sigma$ level are overlaid on the 
HDF I-band image centered on the position of
 HDF850.4 (20\arcsec ~on a side). A 15 $\mu$m detection
 from the complete catalog of
Aussel et al. (1998) has been associated with this
radio source and suggests likely starburst 
activity in the disk galaxy. The symbols are the
same as for Figure 1.
}

\figcaption{Radio 1.4 GHz contours drawn at the  
-2 and 2 $\sigma$ level are overlaid on the 
HDF I-band image in the field of HDF850.5 (20\arcsec
~on a side). This is the one sub-mm source in the HDF
which has no plausible radio counterpart. The symbols
are the same as Figure 1.}

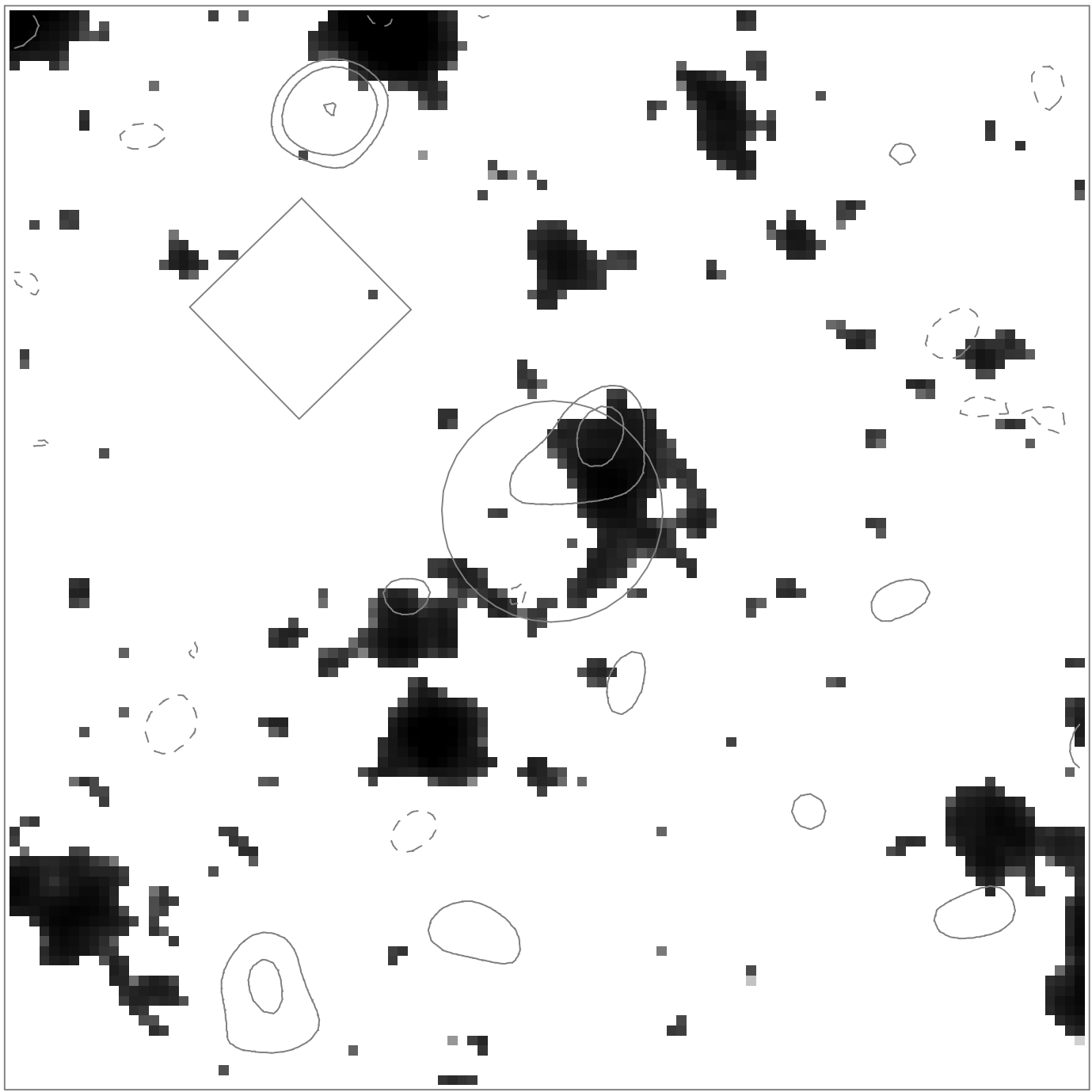
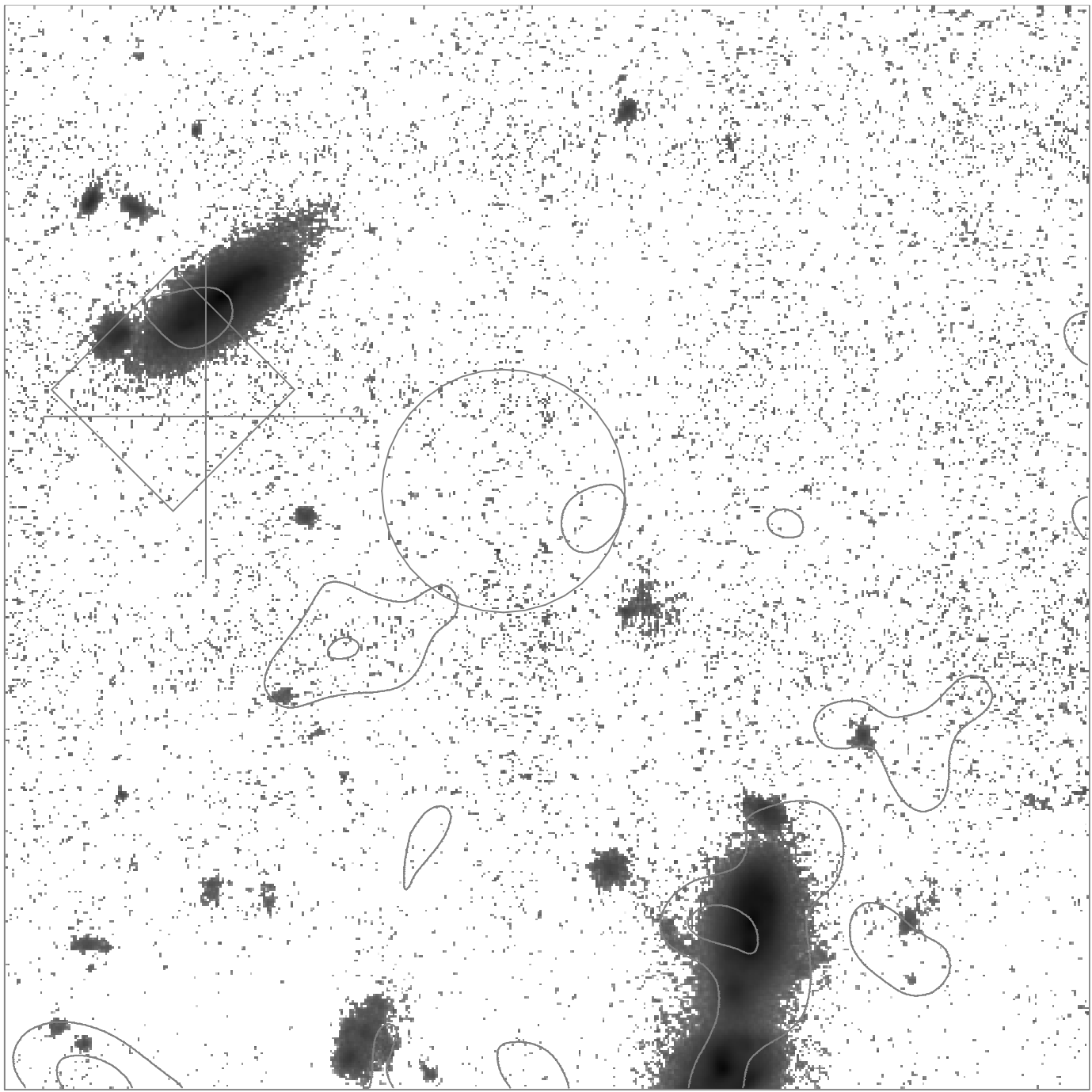
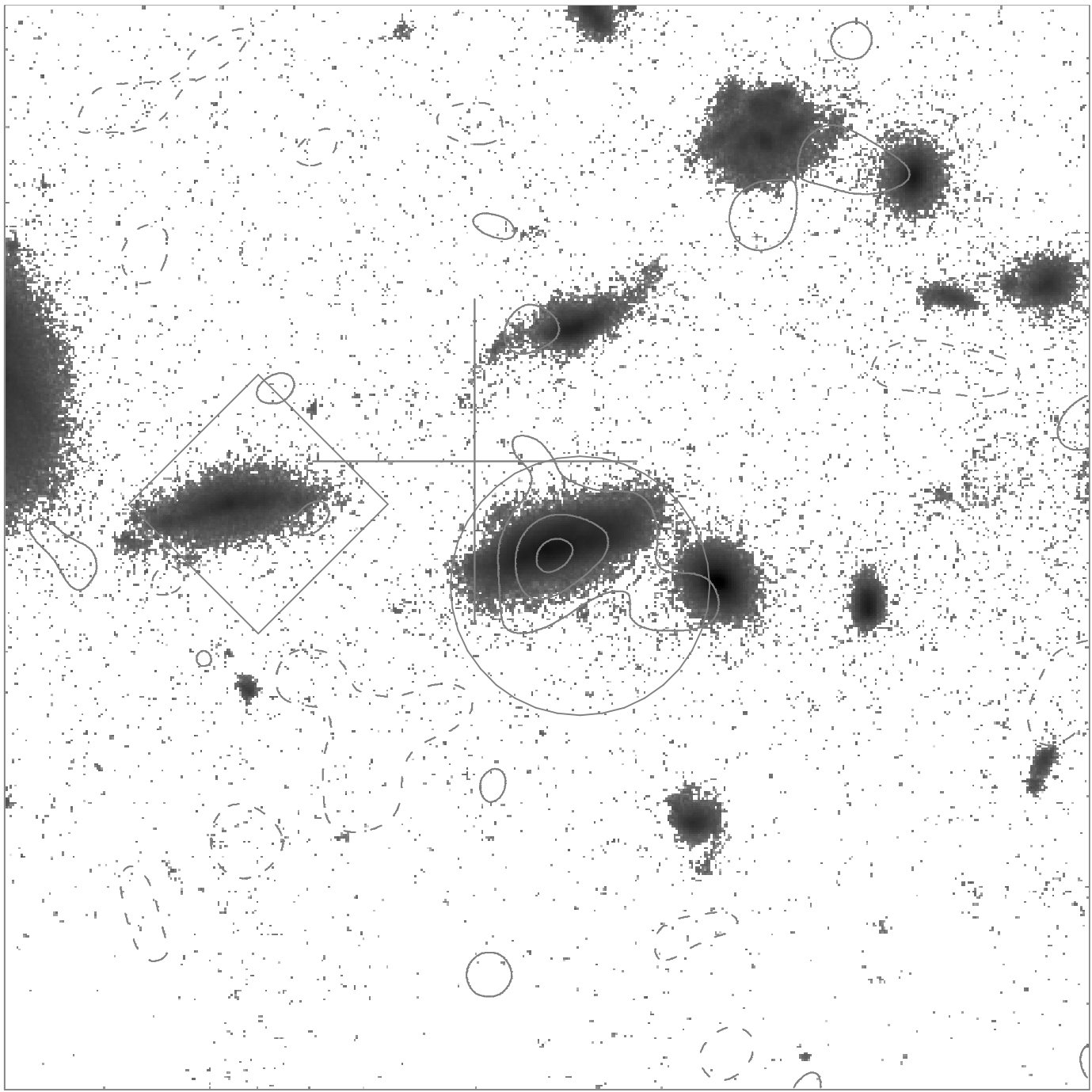
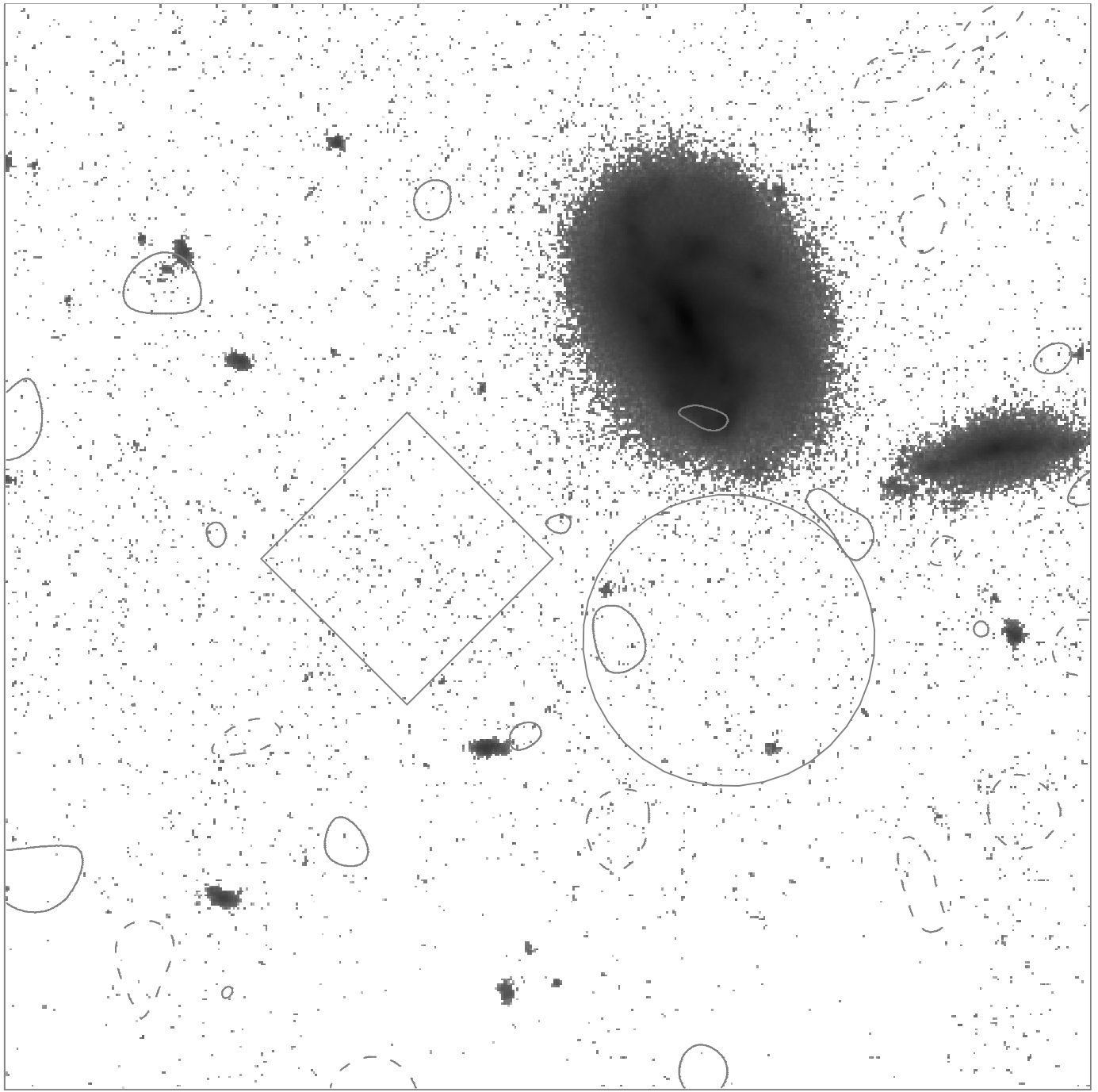
\end{document}